\documentclass[conference]{IEEEtran}
\IEEEoverridecommandlockouts
\usepackage[left=0.625in,right=0.625in,top=0.7in,bottom=1in]{geometry}
\usepackage{cite}
\usepackage{amsmath,amssymb,dsfont,stfloats,color,url,bbm}
\usepackage[pdftex]{graphicx}
\usepackage{subfigure}
\usepackage{mathabx}
\usepackage{multirow}
\usepackage{tabu}
\usepackage{amsthm}
\usepackage{algorithmic}
\usepackage{textcomp}
\usepackage{xcolor}
\usepackage{yfonts}
\usepackage[font=footnotesize]{caption}

\def\BibTeX{{\rm B\kern-.05em{\sc i\kern-.025em b}\kern-.08em
    T\kern-.1667em\lower.7ex\hbox{E}\kern-.125emX}}

\DeclareGraphicsExtensions{.eps,.pdf,.png,.jpg,.gif,.jpeg,.pstex}

\usepackage{mathtools}
\usepackage{cite}
\usepackage{amsmath}
\usepackage{amssymb}
\usepackage{latexsym}
\usepackage{verbatim}
\usepackage{subfigure}
\usepackage{psfrag}
\usepackage{amsthm}
\usepackage{color}
\usepackage{enumitem}

\usepackage{url}              

\usepackage{esdiff}

\usepackage{pgf,tikz,psfrag}
\usepackage{yfonts}

\def\mathlette#1#2{{\mathchoice{\mbox{#1$\displaystyle #2$}}%
		{\mbox{#1$\textstyle #2$}}%
		{\mbox{#1$\scriptstyle #2$}}%
		{\mbox{#1$\scriptscriptstyle #2$}}}}
\newcommand{\matr}[1]{\mathlette{\boldmath}{#1}}

\newfont{\bbb}{msbm10 scaled 700}

\newfont{\bb}{msbm10 scaled 1100}
\newcommand{\CC}{\mbox{\bb C}}

\newcommand{\EE}{\mbox{\bb E}}

\newcommand{\hv}{{\bf h}}

\newcommand{\mv}{{\bf m}}

\newcommand{\rv}{{\bf r}}
\newcommand{\sv}{{\bf s}}

\newcommand{\xv}{{\bf x}}

\newcommand{\zerov}{{\bf 0}}

\newcommand{\Cm}{{\bf C}}

\newcommand{\Id}{{\bf I}}

\newcommand{\Qm}{{\bf Q}}
\newcommand{\Rm}{{\bf R}}
\newcommand{\Sm}{{\bf S}}

\newcommand{\Wm}{{\bf W}}

\newcommand{\Xm}{{\bf X}}
\newcommand{\Ym}{{\bf Y}}
\newcommand{\Zm}{{\bf Z}}

\newcommand{\Ac}{{\cal A}}

\newcommand{\Cc}{{\cal C}}
\newcommand{\Dc}{{\cal D}}

\newcommand{\Nc}{{\cal N}}

\newcommand{\Qc}{{\cal Q}}

\newcommand{\phiv}{\hbox{\boldmath$\phi$}}

\newcommand{\Sigmam}{\hbox{\boldmath$\Sigma$}}

\newcommand{\diag}{{\hbox{diag}}}

\newcommand{\herm}{{\sf H}}

\newcommand{\SNR}{{\sf SNR}}

\usepackage{stmaryrd} 

\usepackage{nohyperref}  

\allowdisplaybreaks

\usepackage{stmaryrd}
\SetSymbolFont{stmry}{bold}{U}{stmry}{m}{n}

\begin{document}

\setlength{\abovedisplayskip}{1pt}
\setlength{\belowdisplayskip}{1pt}
\setlength{\abovedisplayshortskip}{1pt}
\setlength{\belowdisplayshortskip}{1pt}

\title{Joint Message Detection, Channel, and User Position Estimation for 
Unsourced Random Access in Cell-Free Networks}
\author{\IEEEauthorblockN{Eleni Gkiouzepi\IEEEauthorrefmark{1},
		Burak \c{C}akmak\IEEEauthorrefmark{1},
		Manfred Opper\IEEEauthorrefmark{2},
		Giuseppe Caire\IEEEauthorrefmark{1}}
	\thanks{\IEEEauthorrefmark{1} Faculty of Electrical Engineering and Computer Science, Technische Universit\"at Berlin, Germany. (emails: \{burak.cakmak,gkiouzepi,caire\}@tu-berlin.de)}
	\thanks{\IEEEauthorrefmark{2} Faculty of Electrical Engineering and Computer Science,  Technische Universit\"at Berlin, Germany, the Institute of Mathematics, University of Potsdam, Germany and the Centre for Systems Modeling and Quantitative Biomedicine, University of Birmingham, United Kingdom (email: manfred.opper@tu-berlin.de).}
}
\date{\today}

\maketitle
\begin{abstract}
We consider unsourced random access (uRA) in user-centric cell-free (CF) wireless networks, where 
random access users send codewords from a common codebook during specifically dedicated random access channel 
(RACH) slots. The system is conceptually similar to the so-called 2-step RACH currently discussed in 3GPP standardization. 
In order to cope with the distributed and CF nature of the network, we propose to partition the network coverage 
area into zones (referred to as ``locations'') and assign an uRA codebook to each location, such that users in a certain location
make use of the associated codebook. The centralized uRA decoder makes use of the multisource AMP algorithm recently proposed by the authors. 
This yields at once the list of active uRA codewords, an estimate of the corresponding channel vectors, and an estimate of the active users' position. We show excellent performance of this approach
and perfect agreement with the rigorous theoretical ``state evolution'' analysis.  
We also show that the proposed ``location-based'' partitioned codebook approach significantly outperforms
a baseline system with a single non-partitioned uRA codebook. 
\end{abstract}


\section{Introduction}  \label{intro}

\begin{figure}[t]
	\centering	\includegraphics[width=7cm]{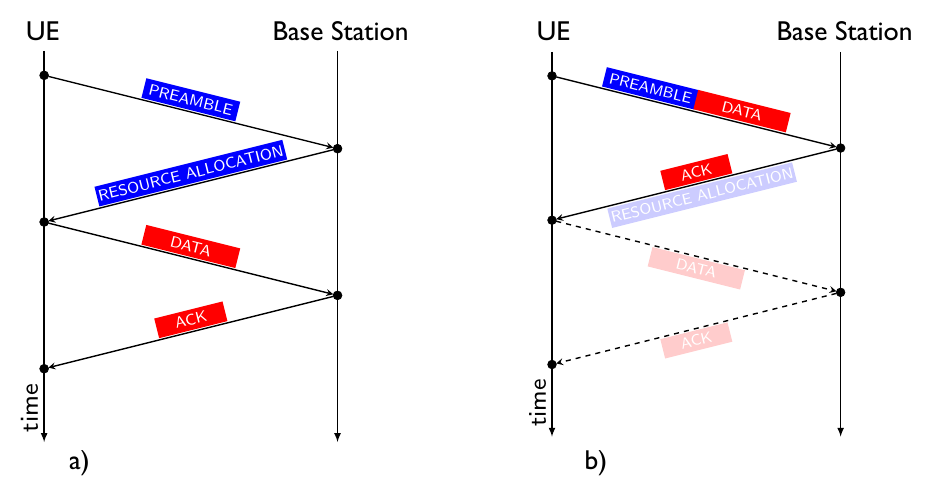}
	\caption{a) Conventional 4-step RACH; b) Novel 2-step RACH.} 
	\label{2step} 
 \vspace{-0.5cm} 
\end{figure}

To support massive connectivity and low
latency, unsourced Random Access (uRA) has been proposed
and widely investigated in several works \cite{polyanskiy2017perspective,fengler2021sparcs,fengler2021non,amalladinne2021unsourced,gkagkos2023fasura}. 
In uRA, a virtually unlimited population of user devices (UEs) make use of the same codebook. 
At every random access (RACH) slot, only a finite number of users is active and transmit a codeword from the uRA codebook. 
These codewords can be seen as ``tokens'', used by the uRA users to be identified and to reserve some transmission opportunity in a subsequent slot. Interestingly, a similar idea is being developed in 3GPP standardization and is referred to as 2-step RACH \cite{peralta2021two}, where the uRA codebook is formed by a collection of ``preamble sequences'', each pointing at a block of  time-frequency referred to as physical uplink shared channel (PUSCH) opportunity (see Fig.~\ref{2step}). 

In a cell-free (CF) user-centric wireless network (e.g., see \cite{demir2021foundations} and references therein), the problem is further complicated by the fact that UEs are spatially distributed.
Since uplink (UL) and downlink (DL) transmission is operated by user-centric clusters of 
radio units (RUs), the system cannot establish such clusters for the uRA users since
their position is a priori unknown. 

In order to cope with the distributed and CF nature of the network, we propose to partition the network coverage 
area into zones (referred to as ``locations'') and assign an uRA codebook to each location, such that users in a certain location
make use of the associated codebook. The centralized uRA decoder makes use of the multisource AMP
algorithm recently proposed by the authors \cite{cakmak2024joint}. 
Messages are detected using a Neyman-Pearson binary hypothesis test on the AMP output. The AMP output also yields the estimates of the channel vectors corresponding to active messages
and an estimate of the active users' position on a quantized grid of positions with much finer resolution with
respect to the coarse locations. As a result, after a RACH slot, the system knows a) the list of active messages; b) the corresponding uplink channel estimates; c) the position of the users sending these messages. 
Hence, the system is able to immediately setup user-centric clusters of RUs for each uRA user, 
receive subsequent uplink data and/or send beamformed downlink data. 
The proposed scheme bears the potential of seamless connectivity
and very-low access latency, by avoiding the lengthy explicit pilot assignment and user-centric cluster
setup phase which is routinely implied (yet rarely analyzed) in the conventional studies on CF user-centric wireless networks. 
\section{System Model}  \label{model}

We consider a CF system where $B$ RUs jointly serve single-antenna users randomly placed in a coverage area $\Dc$. 
RUs are located at points $\nu_b \in \Dc$. Each RU has $M$ antennas. The total number of infrastructure 
antennas is $F = BM$. We partition the coverage area $\Dc$ in $U$ zones $\Dc_u : u \in [U]$, 
referred to as ``locations'', such that $\Dc_u \cap \Dc_{u'} = \emptyset$ $\forall \; u \neq u'$ and $\bigcup_{u\in [U]} \Dc_u = \Dc$. 
Consider an uRA codebook\footnote{We shall refer equivalently to uRA 
``codewords'' or ``messages'', and one should take into account that these are referred to as ``preambles'' in 2-step RACH 3GPP parlance.} where codewords are the columns 
of a matrix $\Sm \in \CC^{L \times N}$. The codebook is partitioned into $U$ subsets
$\Sm = [\Sm_1, \ldots, \Sm_U]$ such that $\Sm_u \in \CC^{L \times N_u}$ with $N = \sum_u N_u$. 
The subcodes are assigned to the locations, such that uRA users accessing the RACH slot 
in location $\Dc_u$ will make use of codewords in $\Sm_u$. This ``location-based'' codebook partition 
establishes a correspondence between the codewords and the large-scale fading coefficients (LSFCs) of the propagation from 
the active uRA users and the RUs. 

Consider a UE in position $q \in \Dc_u$, transmitting codeword $(u,n)$ (i.e., the $n$-th codeword of subcode $\Sm_u$). 
Since the RU positions are fixed, there exist $B$ LSFC functions $\gamma_1(\cdot), \ldots, \gamma_B(\cdot)$ 
such that the LSFC between position $q$ and the $b$-th RU is $g_b = \gamma_b(q)$. 
We assume i.i.d. Rayleigh small-scale fading, such that channel vector between such UE and RU $b$ has i.i.d. 
elements  $\sim \Cc\Nc(0,\gamma_b(q))$.  
The composite channel vector $\hv \in \CC^{1 \times F}$ is obtained by concatenating the 
$1 \times M$  channel vectors of all RU's, such that 
\begin{equation} 
	\hv \sim \Cc\Nc( \zerov, \Sigmam(q)),   \label{gaussian}
\end{equation}
with $\Sigmam(q) = \diag( \gamma_1(q), \ldots, \gamma_B(q) ) \otimes \Id_M$.

The channel vector corresponding to message $(u,n)$ is identically zero if the message is not transmitted (inactive) and 
is Gaussian vector if the message is transmitted (active). 
Let $a_{u,n}$ be a Bernoulli-$\lambda_u$ random variable $\in \{0,1\}$ denoting the activity of message $(u,n)$. 
Hence, the corresponding channel vector conditioned on the UE position $q \in \Dc_u$ is  
$\xv_{u,n} = a_{u,n} \hv_{u,n}$ with $\hv_{u,n}$ distributed as in \eqref{gaussian}. 
Let $\xv_u$ denote a generic channel vector corresponding to codewords in $\Sm_u$. Removing the conditioning with respect to the transmitting UE position, $\xv_u$ has the Bernoulli-Gaussian Mixture density
\begin{equation} 
	p_u(\xv) = (1 - \lambda_u) \delta(\xv)  + \lambda_u \int_{\Dc_u} \mbox{\textswab{g}}(\xv; \zerov, \Sigmam(q)) f_u(q) dq,   \label{zioporco}
\end{equation}
where $f_u(q)$ denotes the probability density function of the user position over location $\Dc_u$ and
$\mbox{\textswab{g}}(\cdot; \mv, \Sigmam)$ denotes the Gaussian 
density function corresponding to $\Cc\Nc(\mv, \Sigmam)$.  The matrix $\Xm_u \in \CC^{N_u \times F}$ containing the 
channel vectors for all (active or inactive) messages at location $\Dc_u$ has i.i.d. rows $\xv_{u,1}, \ldots, \xv_{1,N_u}$ distributed as in \eqref{zioporco}. 

In the following proposed AMP detection/estimation algorithm, we will make use of a discrete approximation of
the distribution of the UE position: for each location $\Dc_u$ we define a suitable (discrete) grid of points $\Qc_u$ such that
we can approximate $p_u(\xv)$ as 
\begin{equation} 
	p_u(\xv) = (1 - \lambda_u) \delta(\xv)  + \lambda_u \sum_{q \in \Qc_u} \mbox{\textswab{g}}(\xv; \zerov, \Sigmam(q)) \delta_u(q),   \label{zioporco1}
\end{equation}
where $\{\delta_u(q) : q \in \Qc_u\}$ is a probability mass function. In particular, for UEs uniformly distributed 
in $\Dc_u$ and a uniform grid $\Qc_u$, we have $\delta_u(q) = 1/|\Qc_u|$ for all grid points $q$.

\subsection{Received signal and SNR normalization}

The signal received at the $F$ RUs antennas over the $L$ symbols of the RACH slot is given 
by the $L \times F$ (time-space) matrix 
\begin{eqnarray} 
	\Ym & = & \sum_{u=1}^U \Sm_u \Xm_u  + \Wm, \label{y_model}
\end{eqnarray}
where $\Wm\sim_{\text{i.i.d.}}\mathcal{CN}(0,\sigma_w^2)$.
We consider a random coding ensemble with codewords $\underline{\sv}_{u,n}\in \CC^{L\times 1} \sim_{\text{i.i.d.}} \Cc\Nc(0,1/L)$. Defining the UL signal-to-noise ratio $\SNR$ as the 
average {\em transmit power} (for a active message) over the noise power, we have the normalization
$\sigma_w^2 = 1/ (L \SNR)$, that is consistent with the SNR ``per-chip'' as defined 
in CDMA (e.g., see \cite{tulino2004random} and references therein, and details in \cite{cakmak2024joint}). 
\subsection{Multisource AMP algorithm}

For message activity detection, channel estimation, and active UEs location estimation, we use
the multisource AMP algorithm given and theoretically analyzed in \cite{cakmak2024joint}. 
For each $u \in[U]$, let $\Xm_u^{(t)}$ denote the AMP estimate of $\Xm_{u}$ at iteration step $t=1,2,\ldots$, with some initialization $\Xm_{u}^{(1)}$ (e.g., the all-zero matrices). The algorithm computes
	\begin{subequations}
		\label{AMP_alg}
		\begin{align}
			\matr \Gamma_u^{(t)}&=\Sm_u \Xm_u^{(t)}-
			{\alpha_u}\Zm^{(t-1)}\Qm_u^{(t)} \\
			\Zm^{(t)}&=\Ym-\sum_{u=1}^{U}\matr \Gamma_u^{(t)}\\
			\Rm_u^{(t)}&=\Sm_u^{\herm}\Zm^{(t)}+{\Xm}_u^{(t)}\\
			\Xm_u^{(t+1)}&=\eta_{u,t}(\Rm_u^{(t)})\  \label{ziopera}
		\end{align}	
	\end{subequations}
with $\Zm^{(0)}=\matr 0$. Here, 
$\eta_{u,t}(\cdot):\CC^{F}\to\CC^{F}$ is a $(u,t)$-dependent {\em denoising function} 
applied row-wise to a matrix argument. 
	The recursive relation for updating $\Qm_u^{(t+1)}$ is given by
	\begin{equation}
		\Qm_u^{(t+1)} = \mathbb{E}[\eta_{u,t}'(\xv_{u}+\matr\phi^{(t)})] \quad \forall t\in[T], \label{onsager}
	\end{equation}
	where $\{\matr \phi^{(t)}\}_{t\in[T]}$ is a zero-mean vector-valued Gaussian process with covariance matrix recusrively computed by the {\em State Evolution} (SE) in \cite[Def.1]{cakmak2024joint} and independent of the 
	random vector $\xv_u \sim p_u(\xv)$ (see \eqref{zioporco1}), 
and $\eta_{u,t}'(\rv)$ denotes the Jacobian matrix of $\eta_{u,t}(\rv)$ (see details in \cite{cakmak2024joint}).
The analysis in \cite{cakmak2024joint} rigorously proves that, as 
$L \rightarrow \infty$ with fixed ratios $N_u/L = \alpha_u$, and finite $F$, 
at any given iteration $t$ the rows of the $\Rm_u^{(t)}$ generated by the AMP algorithms are mutually independent 
and distributed as
\begin{equation} 
\rv_u^{(t)} = \xv_u + \phiv^{(t)}. \label{decoupled}
\end{equation}
We refer to \eqref{decoupled} as the {\em decoupled channel model} relative to the global detection/estimation problem.

In the following, we consider the AMP output $\{\Rm_u^{(T)}, : u \in [U]\}$ after $T$ iterations, 
we let $\widehat{\Xm}_u := \Xm_u^{(T+1)} = \eta_{u,T}(\Rm_u^{(T)})$
and  we let $\phiv^{(T)} \sim \Cc\Nc(\zerov, \Cm^{(T)})$ where $\Cm^{(T)}$ is obtained by $T$ iterations of the SE in \cite[Def.1]{cakmak2024joint}. We also neglect all superscript $T$ for the sake of brevity. 
\subsection{ Denoising function}

Based on the decoupled channel model \eqref{decoupled}, we choose the denoising function
as the Bayesian posterior mean estimator of $\xv_u$ given $\rv_u^{(t)}$, i.e., 
$\eta_{u,t}(\rv_u^{(t)}) = \EE [\xv_u \vert \rv_u^{(t)}]$.
Neglecting the iteration indices $u$ and $t$ for notation simplicity, we derive 
$\eta(\rv)$ and $\eta'(\rv)$ for the observation model $\rv = \xv + \phiv$, where $\xv \sim p_u(\xv)$ given in 
\eqref{zioporco1} and $\phiv \sim \Cc\Nc(\zerov, \Cm)$. 
In the denoiser we use the discrete approximation 
\eqref{zioporco1} even though the UE position has the continuous distribution $f_u(q) : q \in \Dc_u$. 
This mismatch is fully accounted for by the
theory in  \cite{cakmak2024joint}, which applies to a vast family
of sufficiently well-behaved denoising functions.

Consider $a \sim $Bern$(\lambda)$, the random Gaussian-mixture vector
$\hv \sim \sum_{q \in \Qc_u} \mbox{\textswab{g}}(\hv; \zerov, \Sigmam(q)) \delta_u(q)$, and 
let $\eta(\rv) = \mathbb E[a\hv\vert\rv]$  where $\rv = a \hv + \phiv$, where $a, \hv, \phiv$ are mutually independent.  
Then, we have $\eta(\rv) = \EE [ \eta(\rv|q) | \rv ]$,  where $\eta(\rv|q) := \EE[ a\hv |\rv , q]$ takes on the form
\begin{align}
	\eta(\rv|q)
	&=\frac{\mathbb E[\hv\vert\rv,a=1,q]}{1+\Lambda_{\rm map}(\rv|q)}\;\label{etamap}
\end{align}
where $\Lambda_{\rm map}(\rv|q) := \frac{1-\lambda}{\lambda}\frac{p(\rv\vert a=0,q)}{p(\rv\vert a=1,q)}$ is the 
the maximum a posteriori probability (MAP) decision test for $a$ 
when the transmitter location $q$ is known.

Next, using the fact that, $\hv \sim \Cc\Nc(\zerov, \Sigmam(q))$ for given $a = 1$ and $q$, and 
$\rv, \hv, \phiv$ are jointly Gaussian, we have immediately
\begin{align}
	\EE[ \hv | \rv, a = 1,q] 
	& =  \rv (\Sigmam(q) +\Cm)^{-1}\Sigmam(q).  \label{num}
\end{align}
Also, after some algebra, it is not difficult to show that \cite{cakmak2024joint}
\begin{align}
	\Lambda_{\rm map}(\rv|q) &= \frac{1-\lambda}{\lambda} 
	\frac{\vert \Sigmam(q) +\Cm\vert}{\vert \Cm \vert } {\rm e}^{-\rv(\Cm^{-1}-(\Sigmam(q) + \Cm)^{-1})\rv^\herm}.  \label{den}
\end{align}
Finally, removing the conditioning with respect to the transmitter position $q$, yields
\begin{align}
	\eta(\rv) & = \sum_{q \in \Qc_u} \eta(\rv|q) \delta_u(q)
\end{align}
By linearity of differentiation, 
the Jacobian matrix is given by 
\[ \eta'(\rv) = \sum_{q \in \Qc_u} \eta'(\rv|q) \delta_u(q) \]
where the Jacobian matrix $\eta'(\rv|q)$ (with respect to the argument $\rv$),
takes on the form (see  \cite{cakmak2024joint} for details)
\begin{align}
	\eta'(\rv|q)= & \frac{(\Sigmam(q)+\Cm)^{-1}\Sigmam(q)}{1+\Lambda_{\rm map}(\rv|q)} \nonumber \\
	 & +\Lambda_{\rm map}(\rv|q)\Cm^{-1}\eta(\rv|q)^\herm \eta(\rv|q)\;.
\end{align}


\section{Message detection}
\vspace{-0.2cm}
In order to attain a desired balance between the message False Alarm (FA) and Missed Detection (MD) probabilities, 
we use a Neyman-Pearson (NP) likelihood ratio test \cite{poor1998introduction} for each
message $(u,n)$ based on the corresponding $n$-th row of the matrix $\Rm_u$ at the AMP output after $T$ iterations.
Consider a generic such row.  The decoupled channel \eqref{decoupled} yields the
observation model $\rv = a \hv + \phiv$, where the two hypotheses are 
H0 $: a = 0$ and H1 $: a = 1$. The NP test takes on the form
\begin{equation}
	\Lambda_{\rm NP}(\rv) := \frac{p(\rv|a=1)}{p(\rv|a=0)} \underset{a = 0}{\overset{ a=1}{\gtreqless}} \tau, 
	\end{equation}
where $\tau$ is a threshold that parameterizes the optimal tradeoff between FA and MD probabilities. 
Using 
\begin{align}
p(\rv|a = 0) = & \textswab{g}(\rv; \zerov, \Cm) \nonumber \\
p(\rv|a=1) = & \sum_{q \in \Qc_u} \textswab{g}(\rv; \zerov, \Sigmam(q) + \Cm) \delta_u(q), 
\end{align}
we obtain
\begin{align}
	\Lambda_{\rm NP}(\rv) & = \sum_{q\in \Qc_u}  
	\frac{\vert \Cm \vert }{\vert \Sigmam(q) +\Cm\vert} {\rm e}^{\rv(\Cm^{-1}-(\Sigmam(q) + \Cm)^{-1})\rv^\herm}  \delta_u(q).
\end{align}
\section{Active UEs position estimation}

The application of the NP test to all rows $n \in [N_u]$ of all $\Rm_u$ yields a list of 
estimated active messages $\widehat{\Ac}_u$ for each location $\Dc_u$, for all $u\in [U]$. 
Letting $\Ac_u$ the true list of active messages, we let $\Ac_u^{\rm d} = \widehat{\Ac}_u \cap \Ac_u$
and $\Ac_u^{\rm fa} = \widehat{\Ac}_u \setminus \Ac_u$ denote the list of active messages effectively detected as active, and 
non-active messages detected as active (false alarms) for location $\Dc_u$. 

For each message in $\widehat{\Ac}_u$, the system assumes that it was transmitted by some UE in location $\Dc_u$. 
The system is interested in estimating the position of such UE. 
For this purpose,  we consider an {\em approximate} MAP estimator on the position $q \in \Qc_u$, i.e., assuming that the true position is quantized on the grid $\Qc_u$. Under the condition that UE is active and transmits the message $(u,n)$ from the position $q \in \Dc_u$, the corresponding 
row of $\Rm_u$ is distributed as $\rv \sim \textswab{g}(\rv; \zerov, \Sigmam(q) + \Cm)$. 
The posterior distribution of the position given the observation $\rv$ is proportional to 
$\textswab{g}(\rv; \zerov, \Sigmam(q) + \Cm) \delta_u(q)$. This yields the MAP position estimator 
\begin{align} 
\widehat{q} = & \underset{q \in \Qc_u}{\text{arg min}} \left\{ \log\left(\pi^N |\Sigmam(q) + \mathbf{C}|\right) \right . \nonumber \\
 & \left . \;\;\;\;\;\;\;\;\;\;\;\;\;\; + \log \delta_u(q) + \mathbf{r}^\mathsf{H} (\Sigmam(q) + \mathbf{C})^{-1} \mathbf{r} \right\}.
	\label{mle}
\end{align}
To quantify the performance of this position estimator, 
we compare with an ``oracle'' that, for all active user in 
some (unquantized) positions $q_0 \sim f_u(\cdot)$, finds the point on the grid $\Qc_u$ at 
minimum distance, i.e., finds $q^\star =  \underset{q \in \Qc_u}{\text{arg min}}|q - q_0|$. 
We restrict this comparison to messages in  $\Ac_u^{\rm d}$. In fact, position estimation is meaningful  only for the UEs 
that are effectively active and are detected as active. The active users detected as non-active (misdetection events) 
are simply ignored by the system, and they will try to randomly access the channel in some next RACH slot. 
The false alarms correspond to users that actually do not exist, and therefore no subsequent UL or DL 
data transmission will follow anyway.

\section{Channel estimation comparison AMP/genie aided MMSE}  \label{channel-est}

The AMP algorithm also outputs the estimated channel matrices $\{\widehat{\Xm}_u : u \in [U]\}$. 
Again, only the channel vectors (i.e., the rows of $\widehat{\Xm}_u$) corresponding to messages in $\Ac_u^{\rm d}$ 
matter, for the same reason said before. 
In order to evaluate the effectiveness of AMP for channel estimation, 
we compare it with a {\em genie-aided} minimum mean-square error (MMSE) 
estimator that has perfect knowledge of the active messages
$\{\Ac_u\}$ and of the position of the corresponding transmitters. 
The received signal conditioned on the knowledge of the true sets of active messages $\Ac_u$ is given by
\vspace{-0.3cm}
\begin{equation}
	\Ym = \sum_{u = 1}^U \sum_{(u,n)\in \Ac_u} \underline{\sv}_{u,n} \hv_{u,n} + \Wm,  \label{jointly-Gaussian}
\end{equation}
with $\hv_{u,n} \sim  \Cc\Nc(\zerov, \Sigmam(q_{u,n}))$, where $q_{u,n}$ is the position of the active UE transmitting
message $(u,n) \in \Ac$. For known $\{\Ac_u\}$ and $\{q_{u,n} : (u,n) \in \Ac_u\}$, the observation 
$\Ym$ in \eqref{jointly-Gaussian} is jointly Gaussian with any given channel vector. Hence, the optimal estimator 
is the linear MMSE estimator. For given $(u,n) \in \Ac_u$, this linear estimator splits into $B$ separate estimators
for each $1 \times M$ channel vector from the active UE in position $q_{u,n}$ to the $b$-th RU. The generic $b$-th estimator is given by 
\begin{align}
	\widehat{\hv}_{u,n,b} = & \gamma_b(q_{u,n}) \underline{\sv}_{u,n}^{\herm} \Sigmam_b^{-1} \Ym_b, 
	\end{align} 
where $\Ym_b \in \CC^{L \times M}$ is the section of $\Ym$ corresponding to the signal received at RU $b$, and
\[ \Sigmam_b = \left(\sum_{(u',n') \in \Ac} \gamma_b(q_{u',n'}) \underline \sv_{u',n'}\underline \sv_{u',n'}^{\herm}+ \sigma^{2}_{w}\Id_L \right), \]
is the covariance matrix of any column of $\Ym_b$ (conditioned on uRA codebook $\Sm$, 
active messages $\{\Ac_u\}$ and  transmitter positions 
$\{q_{u,n} : (u,n) \in \Ac_u\}$. The resulting MSE per channel coefficient is given by 
\begin{equation}
\frac{1}{M}\EE [ ||\hv_{u,n,b}-\widehat{\hv}_{u,n,b}||^2] = \gamma_b(q_{u,n}) - 
\gamma_b(q_{u,n})^2 \underline{\sv}_{u,n}^\herm \Sigmam_b^{-1} \underline{\sv}_{u,n}.  \label{cond-mmse-genie}
\end{equation}
The unconditional MMSE can be obtained by Monte Carlo averaging 
\eqref{cond-mmse-genie} with respect to 
$\Sm$, $\{\Ac_u\}$, and $\{q_{u,n} : (u,n) \in \Ac_u\}$.

	\vspace{-0.3cm}
\section{Numerical  Results}\label{sim_results}

\begin{figure}[h]
		\vspace{-0.5cm}
	\centering	\includegraphics[width=6cm]{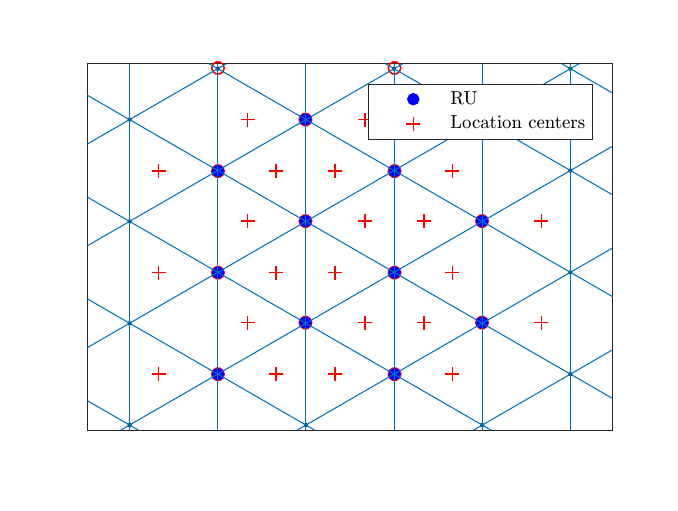}
	\caption{Hexagonal network topology with equilateral triangle tiles used in the numerical results of this paper.} 
	\label{RUs} 
	\vspace{-0.2cm}
\end{figure}

We consider the coverage area $\Dc$ represented in Fig.~\ref{RUs}, with $B=12$ RUs with $M = 2$ antennas each
placed on a hexagonal pattern and $U=24$ location tiles of equilateral triangular shape of side $0.1$ km. Notice that if this was a conventional cellular layout, the locations would correspond to classical 120$^\circ$ sectors. A torus topology is used to avoid border effects. Each location-based  codebook $\Sm_u$ consists of $N_u = 2048$ codewords 
with block length $L = 1024$. The message activity probabilities are set as $\lambda_u \in \{0.009,0.005,0.0005,0.0002\}$, repeated in a fixed raster pattern by columns, bottom to top, left to right.  
This means that the average number of active messages per location ranges between 
$0.0002 \times 2048 \approx 0.4$ to $0.009 \times 2048 \approx 18$. 
The non-uniform message activity probabilities capture the fact that locations may be unequally populated, 
and this aspect is fully taken into account by our scheme. 
The active UE positions are generated independently and uniformly on each location. 

The distance-dependent pathloss (PL) function is given by
\begin{equation}
	\gamma_b(q) =1/\left(1 + \left(|q - \nu_b|/{d_0}\right)^\rho \right)\;,
\end{equation}
with pathloss exponent $\rho = 3.67$, and  3dB cutoff distance 
$d_0=0.01357$ km. To set the UL transmit power (i.e.,  the parameter $\SNR$), we consider the 
distance $\varsigma$ between a location center point (red cross in Fig.~\ref{RUs}) and a nearest RU (blue point in Fig.~\ref{RUs}) and impose a desired meaningful value $\SNR_{\rm rx}$ of the corresponding receiver SNR. This yields
$\SNR  = \SNR_{\rm rx} \times \left(1 + \left(\varsigma/{d_0}\right)^\rho \right)$. In these results we used
$\SNR_{\rm rx} = 10$ dB. 
 
We also compare the proposed location-based codebook partition with  
a scheme that treats the whole CF network as a single ``location'' with a not partitioned uRA codebook
with same total size of $N=49152$ codewords and block length $L=1024$. 
This baseline system is referenced in the following as ``single codebook'' (SC), in contrast with the 
proposed ``location-based'' (LB). 
For the SC system, the message activity probability is $\lambda=0.0037$, yielding the same average number of active messages as the LB system. 
\vspace{-0.4cm}
\begin{figure}[h]
	\centering
		\includegraphics[width=7cm]{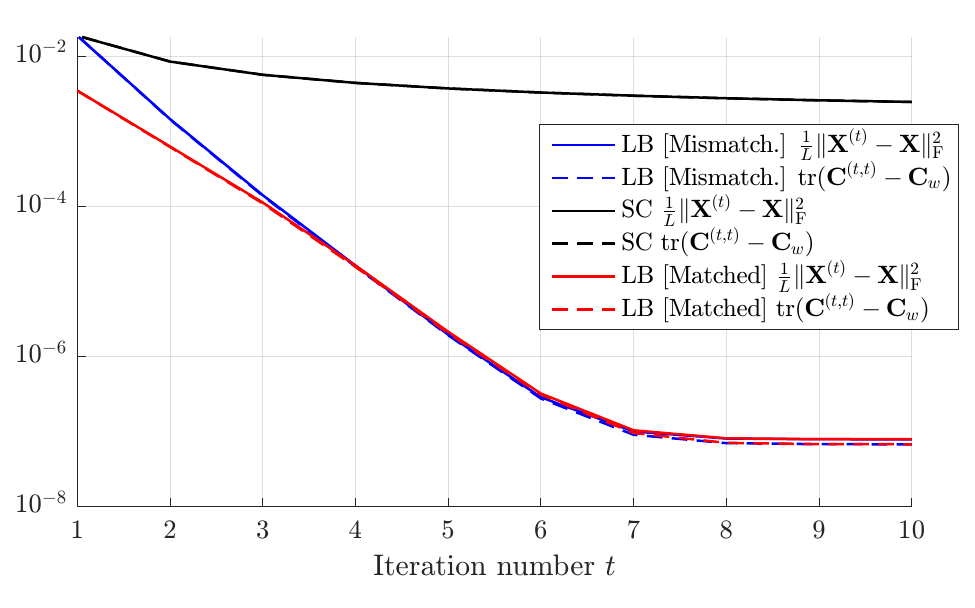}
	\caption{AMP  total normalized MSE and its theoretical deterministic prediction.} 
	\label{fig.SE_Single_vs_Location}    	
	\vspace{-0.1cm}
\end{figure}

In Fig.~\ref{fig.SE_Single_vs_Location} the theory-simulation consistency for the total normalized MSE 
of the AMP iterations $t$ is illustrated. The curves denoted as ``matched'' correspond to the AMP denoising function
given in this paper, for a sufficiently dense quantization grid $\Qc_u$ (see later). 
The curves denoted as ``mismatched'' correspond to using the denoising function that assumes
a ``nominal'' UE position distribution given by a single mass point at the location center (red crosses in Fig.~\ref{RUs}), 
even though the  users are uniformly distributed. We note that this mismatch 
in the posterior mean denoising function yields negligible degradation for the chosen system parameters. 

\begin{figure}[h]
\vspace{-0.2cm}
	\centering
		\includegraphics[width=7cm]{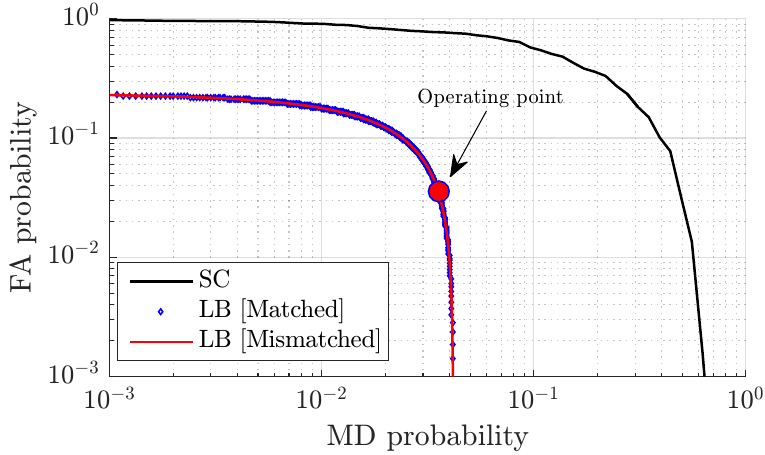}
	\caption{MD and FA probability tradeoff curves for the LB (matched, mismatched) and the SC case.}
	\label{fig.Pfa_Pmd_single_mismatched}    
	\vspace{-0.1cm}	
\end{figure}

Fig.~\ref{fig.Pfa_Pmd_single_mismatched} shows the finite dimensional simulation results 
of the MD and FA probabilities for the same system parameters as before. 
The LB scheme exhibits superior performance compared to the SC baseline. 
For the following results, we operate the system by choosing, for each location $u$, a NP detection 
threshold $\tau_u$ such that the FA and MD probabilities are equal. 
The equivalent point for the average (over all locations) case is shown in Fig.~\ref{fig.Pfa_Pmd_single_mismatched}.
\begin{figure}[h]
	\centering
		\includegraphics[width=7cm]{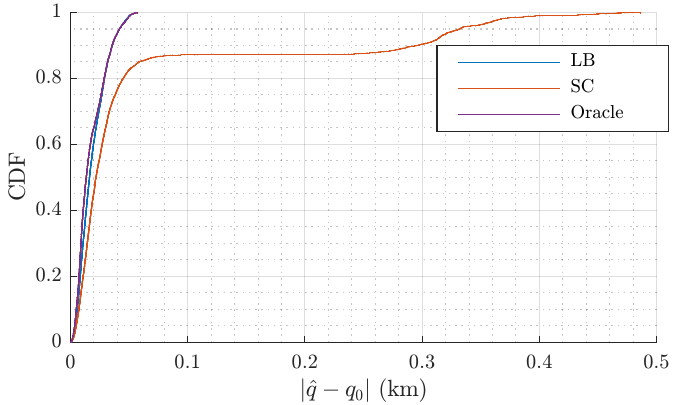}
	\caption{CDF of the absolute position error for the location based and the single codebooks
	schemes, using the proposed MAP position estimator for the active messages.} 
	\label{fig.CDF_Error_position}    	
	\vspace{-0.5cm}	
\end{figure}

Fig.~\ref{fig.CDF_Error_position} shows  the CDF of the absolute error of the position estimation for 
the UEs transmitting messages in $\bigcup_{u \in [U]} \Ac_u^{\rm d}$ (active messages detected as active). 
The MAP estimator in \eqref{mle} can be applied for arbitrary locations and therefore also for the
SC system. We notice that for the location based (LB) partitioned codebook the CDF of the 
position error is almost identical to that obtained by the oracle. We conclude that, with high probability, the position 
estimator finds the point in the grid at minimum distance to the true position. 
In contrast, the SC scheme incurs a much larger position estimation error. 
	\vspace{-0.3cm}	
\begin{figure}[h]
	\centering
	\includegraphics[width=6cm]{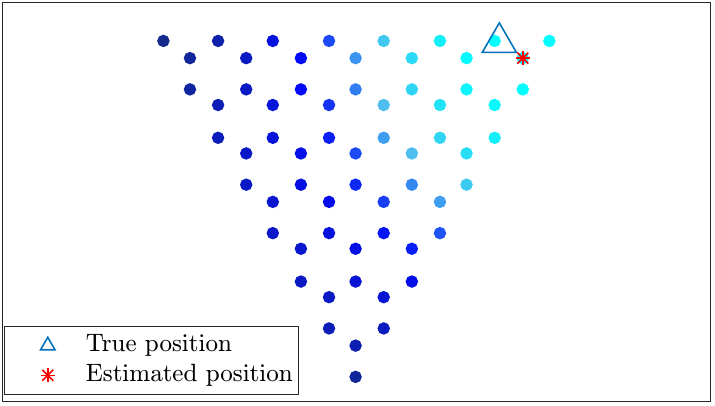}
	\caption{One location $\Dc_u$ snapshot of the proposed MAP position estimator for one active message on a grid of $|\Qc_u|=64$ points.} 
	\label{fig.net_position_est}    
\vspace{-0.25cm}	
\end{figure}

In Fig.~\ref{fig.net_position_est}, one location out of 24 is chosen to demonstrate a snapshot of 
the position estimation. The quantized position grid has $|\Qc_u|=64$ points on a denser hexagonal lattice. 
The color gradient shows the values of MAP position estimator (see eq.\ref{mle}). 
The estimated position $\widehat{q}$ is close to the true position $q_0$. 

\begin{figure}[t]
	\centering
		\includegraphics[width=7cm]{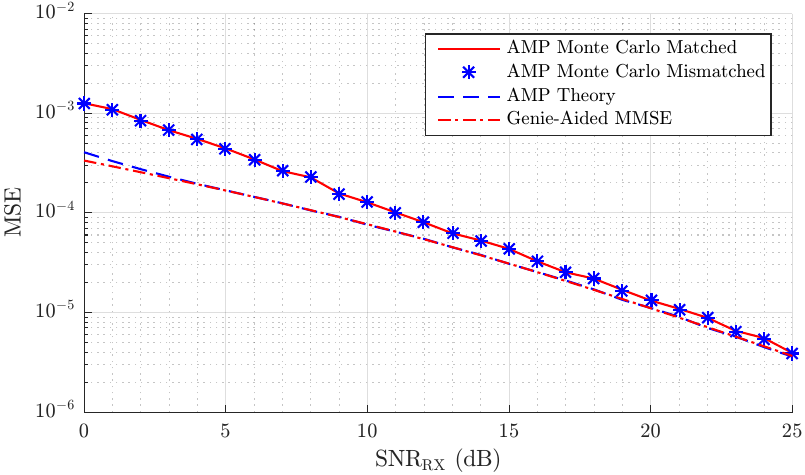}
	\caption{Channel estimation MSE for the active users detected as active for the 
	genie-aided MMSE, the finite dimensional AMP, and the  AMP asymptotic analysis in \cite{cakmak2024joint}.}
	\label{fig.channel_error_genie_vs_AMP_mismatch}  
	\vspace{-0.5cm}	
\end{figure}

Finally, we consider the channel estimation error corresponding to the messages in for the messages in 
$\bigcup_{u \in [U]} \Ac_u^{\rm d}$ (active users detected as active). Fig.\ref{fig.channel_error_genie_vs_AMP_mismatch} shows the average channel estimation MSE where the channel estimate is directly provided by the AMP output (i.e., 
the rows of $\widehat{\Xm}_u$ corresponding to $(u,n) \in \Ac_u^{\rm d}$), compared with the genie-aided MMSE estimator
given in Section \ref{channel-est}. We also show the estimation error provided by the AMP asymptotic analysis in 
 \cite{cakmak2024joint}. For any such estimator, we plot the normalized per-component estimation error 
 as a function of $\SNR_{\rm rx}$ averaged over all locations, over 100 Monte Carlo runs (independent 
 generation of active messages, UE positions, fading channel vectors, and codebook matrices). 
 With the same definitions as in \eqref{cond-mmse-genie}, the MSE per component is defined as
 \[ {\rm MSE} = \frac{1}{FU} \sum_{b=1}^B \sum_{u=1}^U \frac{1}{|\Ac_u^{\rm d}|} \sum_{(u,n) \in \Ac^{\rm d}_u} \EE [ ||\hv_{u,n,b}-\widehat{\hv}_{u,n,b}||^2] \]
for any estimator $\widehat{\hv}_{u,n,b}$. We notice that for the LB system the ``matched'' and ``mismatched''  AMP 
perform almost identically also in terms of channel estimation.  

\section{Conclusion}\label{conclusion}
In this paper, we have implemented the AMP algorithm for uRA in CF user-centric networks, in a centralized manner. The  ``location-based'' model  outperforms the ``single codebook'' approach. The problem of estimating the channel and the unknown user position can be alleviated by using the described  framework, when the LSFCs are random.
\bibliographystyle{IEEEtran}
\bibliography{report,massive-MIMO-references}

\section{Acknowledgements}
The work of Eleni Gkiouzepi, Burak \c{C}akmak, Manfred Opper and Guiseppe Caire was supported by BMBF Germany in the program of ``Souverän. Digital. Vernetzt.'' Joint Project 6G-RIC (Project IDs 16KISK030).

\end{document}